\documentclass[prd,preprint
,superscriptaddress,showpacs,nofootinbib,%
tightenlines]{revtex4}
\usepackage{graphics}
\usepackage{epsfig}
\usepackage{color}
\usepackage{slashed}

\newcommand{\ben}{\begin{displaymath}}
\newcommand{\een}{\end{displaymath}}
\newcommand{\be}{\begin{equation}}
\newcommand{\ee}{\end{equation}}
\newcommand{\bea}{\begin{eqnarray}}
\newcommand{\eea}{\end{eqnarray}}
\begin{document}
\title{The width of the Roper resonance\\ in baryon chiral perturbation theory}
\author{Jambul~Gegelia}
\affiliation{Institute for Advanced Simulation, Institut f\"ur Kernphysik
   and J\"ulich Center for Hadron Physics, Forschungszentrum J\"ulich, D-52425 J\"ulich,
Germany}
\affiliation{Tbilisi State  University,  0186 Tbilisi,
 Georgia}
\author{Ulf-G.~Mei\ss ner}
\affiliation{Helmholtz Institut f\"ur Strahlen- und Kernphysik and Bethe
   Center for Theoretical Physics, Universit\"at Bonn, D-53115 Bonn, Germany}
 \affiliation{Institute for Advanced Simulation, Institut f\"ur Kernphysik
   and J\"ulich Center for Hadron Physics, Forschungszentrum J\"ulich, D-52425 J\"ulich,
Germany}
\author{De-Liang Yao}
 \affiliation{Institute for Advanced Simulation, Institut f\"ur Kernphysik
   and J\"ulich Center for Hadron Physics, Forschungszentrum J\"ulich, D-52425 J\"ulich,
Germany}
\date{June 15, 2016}
\begin{abstract}
   We calculate the width of the Roper resonance at next-to-leading order in a systematic expansion of
baryon chiral perturbation theory  with pions, nucleons, and the delta and Roper resonances as 
dynamical degrees of freedom. 
Three unknown low-energy  constants contribute up to the  given order. 
One of them can be fixed by reproducing the 
empirical  value for the width of the Roper decay into a pion and a nucleon. 
Assuming that the remaining two couplings of the Roper interaction take values 
equal to those of the nucleon, the result for the width of the Roper decaying into a nucleon 
and two pions is  consistent with the  experimental value.

\end{abstract}
\maketitle

\section{Introduction}
\label{Intro}

   At low energies, chiral perturbation 
theory \cite{Weinberg:1979kz,Gasser:1983yg}
provides a successful description of the Goldstone boson sector of QCD.
It turns out that a systematic expansion of loop diagrams in terms of 
small parameters in  effective field theories (EFTs) with
heavy degrees of freedom is a rather complicated issue. 
   The problem of power counting in baryon chiral perturbation 
theory \cite{Gasser:1987rb} may be solved by using the heavy-baryon 
approach \cite{Jenkins:1990jv,Bernard:1992qa,Bernard:1995dp}
or by choosing a suitable renormalization scheme
\cite{Tang:1996ca,Becher:1999he,Gegelia:1999gf,Fuchs:2003qc}.
The $\Delta$ resonance and (axial) vector mesons can  also be included 
in EFT  (see e.g. Refs.~\cite{Hemmert:1997ye,Pascalutsa:2002pi,Bernard:2003xf,Pascalutsa:2006up,Hacker:2005fh,Fuchs:2003sh,Bruns:2004tj,Bruns:2008ub,Terschluesen:2013pya,Leupold:2012qn}).
   On the other hand, the inclusion of heavier baryons such as the Roper 
resonance is more complicated. 

Despite the fact that the Roper resonance was found a long time ago  
in a partial wave  analysis of  pion-nucleon 
scattering data \cite{Roper:1964zza}, a satisfactory theory of this state is still missing. 
The Roper is  particularly interesting as it is the first nucleon resonance 
that exhibits a decay mode into a nucleon and two pions, besides the 
decay into a nucleon and a pion. Also, the Roper appears unexpectedly 
low in the spectrum, below
the first negative parity nucleon resonance, the $S_{11}(1535)$. It is therefore
timely to address this state in a chiral EFT. First steps in this direction 
have been made in Refs.~\cite{Borasoy:2006fk,Djukanovic:2009gt,Long:2011rt,Bauer:2012at,Epelbaum:2015vea}.

In this work we calculate the width of the Roper resonance in a systematic expansion 
in the framework of baryon chiral perturbation theory with pions, nucleons, the delta and 
Roper resonances as explicit degrees of freedom. 
 
The paper is organised as follows: in Section~\ref{EffL}  we specify the effective Lagrangian, 
in Section~\ref{Polemassandw} the pole mass and the width of the Roper 
resonance are defined and  the perturbative calculation of the width is outlined in Section~\ref{sec:width}.
In Section~\ref{Ren} we discuss the renormalization and the power counting applied to 
the decay amplitude of the Roper resonance, while Section~\ref{numerics} contains the  
numerical results. We briefly summarize in Section~\ref{summary}.

\medskip

\section{Effective Lagrangian}
\label{EffL}

    We start by specifying the elements of the chiral effective Lagrangian which are  relevant 
for the calculation of the width of the Roper at next-to-leading order in the power 
counting specified below. We consider  pions,
nucleons, the delta and Roper resonances as dynamical degrees of freedom.
The corresponding most general effective Lagrangian can be written as
\bea
{\cal L}_{\rm eff}={\cal L}_{\pi\pi}+{\cal L}_{\pi N}+{\cal L}_{\pi \Delta}+{\cal L}_{\pi R}
+{\cal L}_{\pi N\Delta}+{\cal L}_{\pi NR}+{\cal L}_{\pi\Delta R},
\eea
where the subscripts indicate the dynamical fields contributing to a given term. 
From the purely mesonic sector we need the following structures \cite{Gasser:1983yg,Bellucci:1994eb}
\bea {\cal
L}_{\pi\pi}^{(2)}&=&  \frac{F^2}{4}\langle\partial_\mu U \partial^\mu U^\dagger\rangle
+\frac{F^2 M^2}{4}\langle U^\dagger+ U\rangle ,\nonumber\\
{\cal L}_{\pi\pi}^{(4)}&=&\frac{1}{8}l_4\langle u^\mu
u_\mu\rangle\langle \chi_+\rangle+\frac{1}{16}(l_3+l_4)\langle
\chi_+\rangle^2, \eea
where $\langle ~~ \rangle$ denotes the trace in flavor space, 
$F$ is the pion decay constant in the chiral 
limit and $M$ is the leading
term in the quark mass expansion of the pion mass 
\cite{Gasser:1983yg}.
   The pion fields are contained in the unimodular unitary $2\times 2$
matrix $U$, with $u=\sqrt{U}$ and 
\begin{eqnarray}
u_\mu & = & i \left[u^\dagger \partial_\mu u -u \partial_\mu u^\dagger
\right],\nonumber\\
\chi^+&=&u^\dag\chi u^\dag+ u\chi^\dag u~,\quad \chi= \left[ \begin{array}{c c}
      M^2 & 0 \\
      0 & M^2 \\
   \end{array}\right].
\label{umudef}
\end{eqnarray}

The terms of the Lagrangian  with pions and baryons contributing to our calculation read:
\bea
{\cal L}_{\pi N}^{(1)}&=&\bar{\Psi}_N\left\{i\slashed{D}-m+\frac{1}{2}g \,\slashed{u}\gamma^5\right\}\Psi_N\, ,\nonumber\\
{\cal L}_{\pi R}^{(1)}&=&\bar{\Psi}_R\left\{i\slashed{D}-m_R+\frac{1}{2}g_R\slashed{u}\gamma^5\right\}\Psi_R\, ,\nonumber\\
{\cal L}_{\pi R}^{(2)}&=&\bar{\Psi}_R\left\{c_1^R\langle\chi^+\rangle\right\}\Psi_R\, ,\nonumber\\
{\cal L}_{\pi NR}^{(1)}&=&\bar{\Psi}_R\left\{\frac{g_{\pi NR}}{2}\gamma^\mu\gamma_5 u_\mu\right\}\Psi_N+ {\rm h.c.}\, ,\nonumber\\
{\cal L}^{(1)}_{\pi\Delta}&=&-\bar{\Psi}_{\mu}^i\xi^{\frac{3}{2}}_{ij}\left\{\left(i\slashed{D}^{jk}-m_\Delta\delta^{jk}\right)g^{\mu\nu}
-i\left(\gamma^\mu D^{\nu,jk}+\gamma^\nu D^{\mu,jk}\right) +i \gamma^\mu\slashed{D}^{jk}\gamma^\nu+m_\Delta\delta^{jk} \gamma^{\mu}\gamma^\nu\right.
\nonumber\\
 &&\left.+\frac{g_1}{2}\slashed{u}^{jk}\gamma_5g^{\mu\nu}+\frac{g_2}{2} (\gamma^\mu u^{\nu,jk}+u^{\nu,jk}\gamma^\mu)\gamma_5+\frac{g_3}{2}\gamma^\mu\slashed{u}^{jk}\gamma_5\gamma^\nu \right\}\xi^{\frac{3}{2}}_{kl}{\Psi}_\nu^l\,  ,\nonumber
 \\
{\cal L}^{(1)}_{\pi N\Delta}&=&h\,\bar{\Psi}_{\mu}^i\xi_{ij}^{\frac{3}{2}}\Theta^{\mu\alpha}(z_1)\ \omega_{\alpha}^j\Psi_N+ {\rm h.c.}\, ,\nonumber\\
{\cal L}^{(1)}_{\pi \Delta R}&=&h_R\,\bar{\Psi}_{\mu}^i\xi_{ij}^{\frac{3}{2}}\Theta^{\mu\alpha}(\tilde{z})\ \omega_{\alpha}^j\Psi_R+ {\rm h.c.}\, ,
\label{LagrNRDP}
\eea
where $\Psi_N$ and $\Psi_R$ are isospin doublet fields
with bare masses $m_{N 0}$ and $m_{R 0}$, corresponding to the nucleon and the Roper resonance, 
respectively.
The vector-spinor isovector-isospinor
Rarita-Schwinger field  $\Psi_\nu$ represents  the $\Delta$ resonance
\cite{Rarita:1941mf} with bare mass $m_{\Delta 0}$,
$\xi^{\frac{3}{2}}$ is the isospin-$3/2$ projector, $\omega_\alpha^i=\frac{1}{2}\,\langle\tau^i u_\alpha \rangle$ and $\Theta^{\mu\alpha}(z)=g^{\mu\alpha}
+z\gamma^\mu\gamma^\nu$, where $z$ is a so-called off-shell parameter. We fix the off-shell structure 
of the interactions involving the delta by adopting $g_1=-g_2=-g_3$ and $z_1=\tilde{z}=0$. 
Note that these off-shell parameters can be absorbed in LECs and are thus redundant \cite{Tang:1996sq,Krebs:2009bf}.
   Leaving out the external sources, the covariant derivatives are defined as follows:
\begin{eqnarray}
D_\mu \Psi_{N/R} & = & \left( \partial_\mu + \Gamma_\mu 
\right) \Psi_{N/R}\,, \nonumber\\
\left(D_\mu\Psi\right)_{\nu,i} & = &
\partial_\mu\Psi_{\nu,i}-2\,i\,\epsilon_{ijk}\Gamma_{\mu,k} \Psi_{\nu,j}+\Gamma_\mu\Psi_{\nu,i}
\,,\nonumber\\
\Gamma_\mu & = &
\frac{1}{2}\,\left[u^\dagger \partial_\mu u +u
\partial_\mu u^\dagger 
\right]=\tau_k\Gamma_{\mu,k}~. \label{cders}
\end{eqnarray}
Note that a mixing kinetic term of the form
$
i\lambda_1\bar{\Psi}_R\gamma_\mu D^\mu\Psi_N-\lambda_2\bar{\Psi}_R\Psi_N+  {\rm h.c}.
$
can be dropped, since, using field transformations and diagonalising the nucleon-Roper mass matrix,  
it can be reduced to the form of operators of the Lagrangian presented above \cite{Borasoy:2006fk}.

\section{The pole mass and the width of the Roper resonance}
\label{Polemassandw}

   The dressed propagator of the Roper resonance can be written as
\begin{equation}
i S_R(p) = \frac{i}{p\hspace{-.45 em}/\hspace{.1em}-m_{R0}
-\Sigma_R(p\hspace{-.45 em}/\hspace{.1em})}\,,\label{dressedDpr}
\end{equation}
where $-i\,\Sigma_R (p\hspace{-.45 em}/\hspace{.1em})$ is the self-energy, i.e. the sum
of all one-particle-irreducible diagrams contributing to the
two-point function of the Roper resonance. 
   The pole of the dressed propagator $S_R$ is obtained by solving the equation
\begin{equation}
S_R^{-1}(z) \equiv z - m_{R0} -\Sigma_R(z)=0\,. \label{poleequation}
\end{equation}
   We define the physical mass and the width of the Roper resonance by parameterizing 
the pole as  
\begin{equation}
z = m_R -i\,\frac{\Gamma_R}{2} \,.\label{poleparameterized}
\end{equation}

\begin{figure}[htbp]
\begin{center}
\epsfig{file=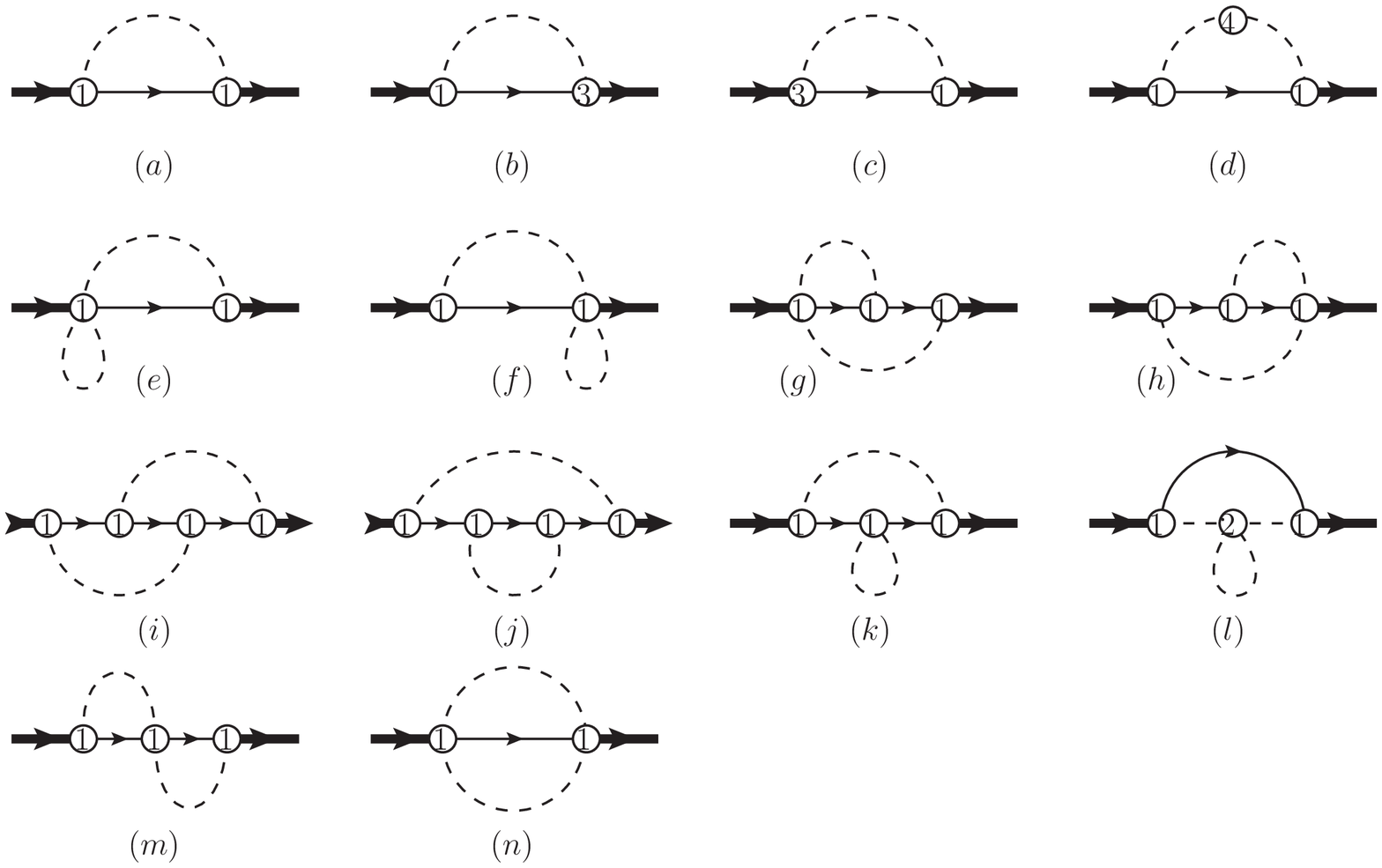,scale=0.55}
\caption{One and two-loop self-energy diagrams of the Roper resonance up-to-and-including 
fifth order according to the standard power counting. The dashed and thick solid lines 
represent the pions and the Roper resonances, respectively. The thin solid lines in the 
loops stand for either nucleons, Roper or delta resonances. The numbers in the circles give
the chiral order of the vertices.}
\label{figSE2}
\end{center}
\end{figure}

   The pertinent topologies of the one- and two-loop diagrams contributing to the 
self-energy of the Roper resonance are shown in Fig.~\ref{figSE2}. 
We use  BPHZ renormalization by subtracting loop diagrams in their chiral limit and replace 
the parameters of the Lagrangian by their renormalized values, i.e. counterterm diagrams are 
not shown explicitly.

We solve Eq.~(\ref{poleequation}) perturbatively order by order in the loop expansion. 
We parameterize the pole as
\begin{equation}
z=m_{2}+\hbar \delta z_1+\hbar^2 \delta z_2 +{\cal O}(\hbar^3),
\label{zpert}
\end{equation}
where $m_2=m_R^0+4 c_1^R M^2 $, with $m_R^0$ the physical Roper mass in the chiral limit, and 
substitute in Eq.~(\ref{poleequation}) in which we write the self-energy as an 
expansion in the number of loops 
\begin{equation}
\Sigma_R = \hbar \Sigma_1+\hbar^2 \Sigma_2 +{\cal O}(\hbar^3)\, .
\label{SEpert}
\end{equation}
By expanding in powers of $\hbar$, we get
\begin{equation}
\hbar \delta z_1+\hbar^2 \delta z_2-\hbar \Sigma_1(m_2) -\hbar^2 \delta z_1 \Sigma_1'(m_{2})
-\hbar^2 \Sigma_2(m_{2}) +{\cal O}(\hbar^3) =0\,. \label{poleequationpert}
\end{equation}
Solving Eq.~(\ref{poleequationpert}) we obtain
\begin{eqnarray}
\delta z_1 & = & \Sigma_1(m_2),\nonumber\\
\delta z_2 & = & \Sigma_1(m_2) \,\Sigma_1'(m_{2})+\Sigma_2(m_{2}).
\label{solvezpert}
\end{eqnarray}
Eq.~(\ref{solvezpert}) leads to the following expression for the width
\begin{eqnarray}
\Gamma_R &=& \hbar \ 2 i \,{\rm Im} \left[\Sigma_1(m_2)\right] \nonumber\\
&+& \hbar^2 \ 2 i \, \biggl\{
{\rm Im} \left[\Sigma_1(m_2)\right]{\rm Re} \left[\Sigma_1'(m_2)\right]
+ {\rm Re} \left[\Sigma_1(m_2)\right]{\rm Im} \left[\Sigma_1'(m_2)\right] \biggr\} \nonumber\\
&+& \hbar^2 \ 2 i \, {\rm Im} \left[\Sigma_2(m_2)\right] +{\cal O}(\hbar^3).
\label{widthpert}
\end{eqnarray}
Using the power counting specified in section~\ref{Ren}, it turns out that the 
contribution of the second term in Eq.~(\ref{widthpert}) is of an order higher than 
the accuracy of our  calculation, which is $\delta^5$ (where $\delta$ is a small expansion 
parameter).  In particular, ${\rm Im} \left[\Sigma_1(m_2)\right]$ is of  order 
$\delta^3$, ${\rm Re} \left[\Sigma_1'(m_2)\right]$ is of order $\delta^4$, 
${\rm Re} \left[\Sigma_1(m_2)\right]$ is of  order $\delta^6$  and 
${\rm Im} \left[\Sigma_1'(m_2)\right]$ is of  order $\delta^2$.
Also, modulo higher order corrections, we can replace $m_2$ by the physical mass $m_R$.  
To calculate the contributions of the one- and two-loop self-energy diagrams to
the width of the Roper resonance, specified in the first and third terms of  
Eq.~(\ref{widthpert}), respectively,  we use the Cutkosky cutting rules.
   As shown in Ref.~\cite{Veltman:1963th}
in quantum field theories with unstable particles the scattering amplitude is unitary 
in the space of stable particles alone. 
Thus, to calculate the imaginary part of the self-energy of the Roper resonance at one 
loop order we need to take into account only the contribution of the diagrams with 
internal nucleon lines. 
At two-loop order only contributions obtained by cutting the lines, corresponding to stable 
particles, are needed.  Details of the calculation of the Roper resonance width using the decay 
amplitudes are given in the next section.

\section{The width of the Roper resonance obtained from the decay amplitudes}
\label{sec:width}

\begin{figure}[htbp]
\begin{center}
\epsfig{file=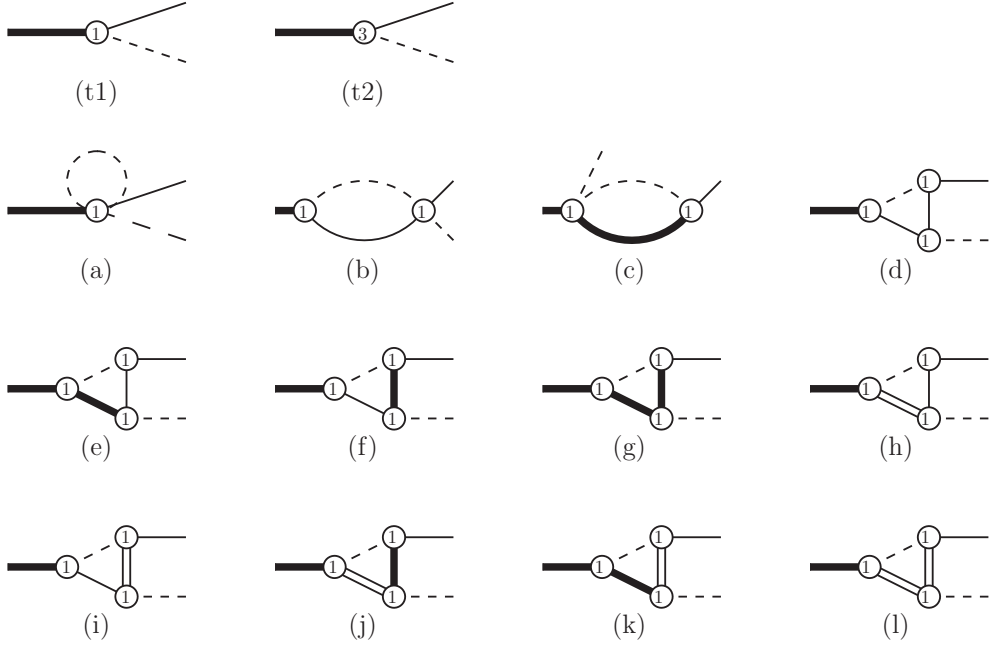,scale=0.7}
\caption{Feynman diagrams contribution to the decay $R\to N\pi$ up to 
leading one-loop order. Dashed, solid, double and thick solid lines 
correspond to pions, nucleons, deltas and Roper resonances, respectively. 
The numbers in the circles give the chiral orders of the vertices.}
\label{treePiND}
\end{center}
\end{figure}

\begin{figure}[t]
\begin{center}
\epsfig{file=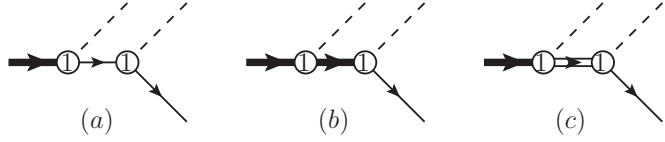,scale=0.7}
\caption{Tree diagrams contributing to the $R\to \pi\pi N$ decay. Crossed diagrams are 
not shown. Dashed, solid, double and thick solid lines 
correspond to pions, nucleons, deltas and Roper resonances, respectively. The numbers in 
the circles give the chiral orders of the vertices.}
\label{figRoper_pipiN}
\end{center}
\end{figure}

By applying the cutting rules to the diagrams of Fig.~\ref{figSE2} 
we obtain the graphs contributing in the decay amplitudes 
of the Roper resonance into $\pi N$ and $\pi \pi N$ systems, specified in 
Figs.~\ref{treePiND} and \ref{figRoper_pipiN}, respectively. 
The decay amplitude corresponding to $R(p)\to N(p^\prime)\pi^a(q)$ 
can be written as
\bea
{\cal A}^{a}=\bar{u}_N(p')\left\{ {A}\,\slashed{q}\gamma_5\tau^a\right\}u_R(p)\, ,
\label{RpiNampl}
\eea
where $a$ is an isospin index of the pion, and the $\bar u, u$ are conventional spinors. 
The corresponding decay width reads
\bea
\Gamma_{R\to\pi N}=\frac{\lambda^{1/2}(m_R^2,m_N^2,M^2)}{16\pi\,m_R^3}\left|{\cal M}\right|^2,
\label{piNDW}
\eea
with  $\lambda(x,y,z)=(x-y-z)^2-4 y z$ and the unpolarized squared amplitude has the form
\bea
\left|{\cal M}\right|^2&=&
3(m_N+m_R)^2\left[(m_N-m_R)^2-M_\pi^2\right]A^\ast A\, .
\label{widthpiN}
\eea  

Next, we define the kinematical variables for the decay $R(p)\to N(p^\prime)\pi^a(q_1)\pi^b(q_2)$   via
\bea
&&s_1=(q_1+q_2)^2\, ,\qquad s_2=(p^\prime+q_1)^2\, ,\qquad s_3=(p^\prime+q_2)^2\, ,
\eea
subject to  the contraint
\bea
s_1+s_2+s_3=m_R^2+m_N^2+2M_\pi^2 ~.
\eea
The isospin and the Lorentz decomposition of the decay amplitude reads
\bea
{\cal A}^{ab}&=&\chi_N^\dagger\left\{\delta^{ab}F_++{i}\epsilon^{abc}\tau^cF_-\right\}\chi_R^{}\ ,\\
F_{\pm}&=&\bar{u}_N(p^\prime)\left\{F_{\pm}^{(1)}-\frac{1}{2(m_N+m_R)}
\left[\slashed{q}_1,\slashed{q}_2\right] F_{\pm}^{(2)}\right\}u_R(p)\, ,
\eea
with the $\chi$ being isospinors,  $a$ and $b$ are isospin indices of the pions. 
The unpolarized squared invariant amplitude is given by
\bea
|{\cal M}|^2
&=& \sum_{i,j=1}^2{\cal Y}_{ij}\left[\frac{3}{2}\,{F_{+}^{(i)}}^\ast F_{+}^{(j)}+3\,{F_{-}^{(i)}}^\ast F_{-}^{(j)}\right] ,\nonumber \\
{\cal Y}_{11}&=&2\left[(m_N+m_R)^2-s_1\right],\nonumber\\
{\cal Y}_{12}&=&{\cal Y}_{21}=-s_1 \nu \,,\nonumber\\
{\cal Y}_{22}&=&\frac{1}{2}\left[(4 M_\pi^2-s_1)(s_1-(m_R-m_N)^2)-s_1\nu^2\right],
\label{Msquared}
\eea
with $\nu$ given by
\be
\nu=\frac{s_2-s_3}{m_N+m_R}\,.
\label{defnu}
\ee 
The decay width corresponding to the $\pi\pi N$ final state is obtained by 
substituting $|{\cal M}|^2$ from Eq.~(\ref{Msquared}) in the following formula
\bea
\Gamma_{R\to \pi\pi N}=\frac{1}{32m_R^3(2\pi)^3}\int_{4M_\pi^2}^{(m_R-m_N)^2}{\rm d}s_1
\int_{s_{2-}}^{s_{2+}}{\rm d}s_2\,|{\cal M}|^2\, ,
\label{defGammapipiN}
\eea
where the integration limits over $s_2$ are given by
\bea
s_{2\pm}=\frac{m_R^2+m_N^2+2M_\pi^2-s_1}{2}\pm\frac{1}{2s_1}\lambda^{1/2}(s_1,m_R^2,m_N^2)
\lambda^{1/2}(s_1,M_\pi^2,M_\pi^2)\, .
\eea

Let us emphasize here that to obtain the width of the Roper resonance we need to calculate 
the imaginary part of the self-energy in the complex region.
However, within the accuracy of our calculation, we only need the imaginary parts of 
the one- and two-loop diagrams for the real mass of the Roper resonance. We can
relate these to the decay amplitudes also calculated by putting the Roper external 
line on the real mass-shell.  
While this is an useful approximation well suited for our current accuracy, to define 
the physical properties of unstable particles one needs to use the 
complex on-shell conditions, see, e.g. Ref.~\cite{Gegelia:2009py}.

Thus, the contributions of the one- and two-loop self-energy diagrams in the width of the Roper 
resonance, specified in the first and third terms of  Eq.~(\ref{widthpert}) sum up to
\begin{equation}
\Gamma_R = \Gamma_{R\to \pi N} + \Gamma_{R\to \pi\pi N},
\label{finalGamma}
\end{equation}
where $\Gamma_{R\to \pi N}$ and $\Gamma_{R\to \pi\pi N}$ are given by Eq.~(\ref{piNDW}) and 
Eq.~(\ref{defGammapipiN}), respectively.
The power counting and the diagrams contributing to each of these decay modes up 
to a given order of accuracy are discussed in the next section.

\section{Renormalization and  power counting}
\label{Ren}

By counting the mass differences $m_R-m_N$, $m_\Delta-m_N$ and $m_R-m_\Delta$ as of the 
same order as the pion mass and the pion momenta, the standard power counting  would apply 
to all tree and loop diagrams considered in this work. 
According to the rules of this counting a four-dimensional loop integration is 
of order $q^4$,  an interaction vertex obtained from
an ${\cal O}(q^n)$ Lagrangian counts as of order $q^n$, a pion
propagator as order $q^{-2}$, and a nucleon propagator as order
$q^{-1}$. We would also assign the order $q^{-1}$ to the
$\Delta$ and the Roper resonance propagators  for non-resonant kinematics. The propagators of 
the delta and the Roper  resonance get enhanced for resonant kinematics when they 
appear as intermediate states outside the loop integration. In this case we would assign 
the order $q^{-3}$ to these propagators. 

As the mass difference $m_R-m_N\sim 400$ MeV, the above mentioned power counting 
cannot be trusted. By considering  $m_R-m_N$ as a small parameter of the 
order $\delta^1$, it is  more appropriate to count $M_\pi\sim \delta^2$.
To work out further details of the new counting, it is  a more convenient
to work with the  kinematical variable $\nu$ as defined in Eq.~(\ref{defnu})
for the $R\to\pi\pi N$ decay.
Within the range of integration specified by  Eq.~(\ref{defGammapipiN}), 
$\nu$ varies from $m_N-m_R$ to $m_R-m_N$  (for $M_\pi=0$) and therefore 
we count $\nu\sim\delta$. As $s_1$ varies from $4 M_\pi^2$ to $(m_R-m_N)^2$, we assign 
the order $\delta^2$ to it. We also count  $m_R-m_\Delta\sim\delta^2$.  

The $R\to\pi N$ width of Eq.~(\ref{widthpiN}) is of order  $\delta^3 \times 
{\rm order \ of \ } A^* A$. The tree and one loop diagrams, contributing to the $R\to \pi N$ 
decay are shown in Fig.~\ref{treePiND}. The tree order diagram (t2) is proportional to 
$\slashed q\gamma_5 M_\pi^2$ and therefore 
it contributes at order $\delta^4$ to $A$, while diagram (t1) gives an order 
$\delta^0$ contribution. All one loop diagrams are of the order $q^3$ in the standard counting. 
Thus they give order $q^2$ contributions in $A$.  Expanding these contributions in powers of 
$M_\pi$, we absorb the first, $M_\pi$-independent, term in the renormalization of the coupling 
of the tree diagram (t1), which becomes complex. However, the imaginary part is of the order 
$\delta^2$ and can be calculated explicitly. 
The next term in the expansion in powers of $M_\pi$ is linear in $M_\pi$ and hence, if non-vanishing, 
it does not violate the standard power counting (terms, non-analytic in $M_\pi^2$ do not violate 
the standard power counting). Therefore its coefficient must contain at least one power of $(m_R-m_N)$. That is, the term linear in $M_\pi$ is at least of order $\delta^3$. Further terms are of even 
higher order. As a result, restricting ourselves to the order $\delta^2$ in $A$, and thus to 
order $\delta^5$ in the width of the Roper resonance, the only contribution of one 
loop diagrams of Fig.~\ref{treePiND} which we might need is the $M_\pi$-independent imaginary part. 
However, this imaginary part starts contributing in $A^*A$ only at order $\delta^4$. Thus all 
contributions of the one-loop diagrams are beyond the accuracy of our calculation. 
We have checked that for the numerical values of the couplings, as specified below, the individual 
contributions of the diagrams in Fig.~\ref{treePiND} in the decay amplitude are 
indeed small compared to the one of the tree order diagram.

According to Eq.~(\ref{defGammapipiN}) the $R\to\pi\pi N$ width is of order  $\delta^3 \times 
{\rm order \ of \ } |{\cal M}|^2 $.   
The leading order tree diagrams contributing to the $R\to\pi\pi N$ decay are shown in 
Fig.~\ref{figRoper_pipiN}. 
The delta propagators in these diagrams are to be understood as dressed ones. Expanding these 
propagators around their pole, we observe that the non-pole parts start contributing 
at higher orders and therefore can be dropped. 
The contributions of the loop diagrams are suppressed by additional powers of $\delta$ 
so that they do not contribute at order  $\delta^5$. 
Among these loop diagrams are those contributing to the decay of the Roper resonance 
to $\pi \pi N$ system with two final pions in the iso-singlet channel. Due to the presence 
of an iso-scalar scalar resonance $f_0(500)$ in this channel \cite{Agashe:2014kda} 
an infinite number of pion-pion finite state interaction diagrams have to be summed up 
(see, e.g., Refs.~\cite{Black:2000qq,Hernandez:2002xk}). 
Alternatively one can include the $f_0(500)$ as an explicit degree of freedom in the 
effective Lagrangian \cite{Meissner:1999vr}. In both approaches it turns out that the 
corresponding contributions to $R\to \pi\pi N$ amplitude are of higher order than $\delta^5$, 
and hence estimated to be within the theoretical uncertainty due to higher order contributions
given in  Eq.~(\ref{RtoNpipi}) below.

\section{Numerical results}
\label{numerics}

To calculate the full decay width of the Roper resonance we use the following standard 
values of the parameters \cite{Agashe:2014kda} 
\begin{eqnarray}
&& M_\pi = 139 \ {\rm MeV}, \ \ m_N=939 \ {\rm MeV}, \  \  m_\Delta=1210\ {\rm MeV}, \  \  
\Gamma_\Delta=100\ {\rm MeV},  \nonumber\\ 
&& \ \  m_R=1365 \ {\rm MeV},  F_\pi=92.2 \ {\rm MeV}, 
\label{Nvalues}
\end{eqnarray}
in Eqs.~(\ref{piNDW}) and (\ref{defGammapipiN}) and obtain 
\begin{eqnarray}
\Gamma_{R\to\pi N}&=&  550 \, g_{\pi NR}^2\ {\rm MeV},\nonumber\\
\Gamma_{R\to\pi\pi N}&=& (1.49\,g_A^2 \,g_{\pi NR}^2-2.76\, g_A^{} \, g_{\pi NR}^2\,g_R^{}+1.48\,g_{\pi NR}^2\, g_R^2\nonumber\\
&+&2.96\,g_A^{}\, g_{\pi NR}^{} \,h^{} h_R^{} -3.79\,g_{\pi NR}^{}\,g_R^{} \,h^{} h_R^{} +9.93\,h^2h_R^2) \ {\rm MeV}.
\label{GammaF}
\end{eqnarray}
Eq.~(\ref{GammaF}) depends on five couplings in total for the two of which we 
substitute $g_A=1.27$ \cite{Agashe:2014kda} and 
$h = 1.42\pm 0.02$. The latter value is the real part of this coupling taken 
from Ref.~\cite{Yao:2016vbz}. As noted before, the imaginary part only contributes
to orders beyond the accuracy of our calculations. As for the other unknown parameters, 
we choose to pin down $g_{\pi NR}$ so that we reproduce the width 
$\Gamma_{R\to \pi N}=(123.5\pm 19.0)$~MeV from PDG~\cite{Agashe:2014kda}, 
which yields $g_{\pi NR}=\pm (0.47\pm 0.04)$. In what follows we take both signs into 
account which contributes to the error budget.  Further we assume $g_R=g_A$ and $h_{R}=h$.
With the values specified above, one can predict the decay width for the 
decay mode $R\to \pi\pi N$:
\bea
\Gamma_{R\to\pi\pi N}&=& \left[0.53(9)-0.98(17)+0.53(9)\pm 3.57(31)\mp 4.57(40)+40.4(1.6)\right]\ 
{\rm MeV}\nonumber\\ 
&=&[40.5(1.6)\pm 1.0(0.5)]~{\rm MeV}~,
\label{Gammanumbers}
\eea
where the second term is due to the choice of the sign of $g_{\pi NR}$. If we 
incorporate the second term as error to the first term, then the decay width reads
\be\label{Gammafinal}
\Gamma_{R\to\pi\pi N}=40.5(2.2)~{\rm MeV}~,
\ee
where the error is obtained  in quadrature from Eq.~(\ref{Gammanumbers}). As is 
clearly seen from Eq.~(\ref{Gammanumbers}), the largest contribution in $\Gamma_{R\to\pi\pi N}$ 
width comes from the decay with the delta resonance as an intermediate state. Further, 
we estimate the theoretical error due to the omitting the higher order contributions 
using the approach 
of Ref.~\cite{Epelbaum:2014efa}, which leads to
\bea\label{RtoNpipi}
\Gamma_{R\to\pi \pi N}&=&(40.5\pm 2.2\pm 16.8) \ {\rm MeV}.
\eea
Our estimation is consistent with $\Gamma_{\pi\pi N}=(66.5\pm 9.5)$ MeV 
quoted by PDG \cite{Agashe:2014kda}.

\section{Summary}
\label{summary}

\medskip

   In current work we have  calculated the width of the Roper resonance up to next-to-leading 
order in a systematic expansion of baryon chiral perturbation theory with pions, nucleons, delta 
and Roper resonances as dynamical degrees of freedom. We define the physical mass and 
the width of the Roper resonance by relating them to the real and imaginary parts of 
the complex pole of the dressed propagator. The next-to-leading order calculation of the width 
requires obtaining the imaginary parts of one- and two-loop self-energy diagrams.  
We employed the Cutkosky cutting rules and obtained the width up to given order 
accuracy by squaring the decay amplitudes. Three unknown coupling constants 
contribute in the corresponding expressions. One of them we fix by reproducing the 
PDG value for the width of the Roper decay in a pion and a nucleon. Assuming that 
the remaining two couplings of the Roper interaction take values equal to those of 
the nucleon, we obtain the result for the width of Roper resonance decaying in the two pions and 
a nucleon that is consistent with the  PDG value, based on the  the uncertainties within the given and
from higher orders, cf. Eq.~(\ref{RtoNpipi}).

To improve the accuracy of our calculation, three-loop contributions to the 
self-energy of the Roper resonance need to be calculated. Moreover, contributions of 
an infinite number of diagrams, corresponding to the scalar-isoscalar pion-pion scattering 
need to be re-summed either by solving pion-pion scattering equations or including the 
$f_0(500)$ as an explicit dynamical degree of freedom. Note also that new unknown 
low-energy coupling constants appear at higher orders, which need to be pinned down.

\acknowledgments
This work was supported in part by Georgian Shota Rustaveli National
Science Foundation (grant FR/417/6-100/14) and by the DFG (TR~16 and CRC~110).
The work of UGM was also supported by the Chinese Academy of Sciences (CAS) President’s
International Fellowship Initiative (PIFI) (Grant No. 2015VMA076).

\end{document}